\documentclass[aps,amsmath,twocolumn]{revtex4}
\usepackage{graphics, setspace,epsfig,color,subfigure}
\usepackage[letterpaper, dvips,width=7.5in,height=8.5in,includemp=false]{geometry}
\usepackage[vcentermath]{}

\newcommand{\be}{\begin{equation}}
\newcommand{\ee}{\end{equation}}

\begin{document}


\title[title]{Strange results from chiral soliton models}
\author{Aleksey Cherman and Thomas D. Cohen }
\affiliation{Department of Physics \\
University of Maryland\\College Park, MD 20742}

\begin{abstract}
The standard collective quantization treatment of the strangeness
content of the nucleon in chiral soliton models such as the
Skyrmion is shown to be inconsistent with the semi-classical
expansion on which the treatment is based.  The strangeness
content vanishes at leading order in the semi-classical expansion.
Collective quantization correctly describes some contributions to
the strangeness content  at the first nonvanishing order in the
expansion, but neglects others at the same order---namely, those
associated with continuum modes. Moreover,  there are fundamental
difficulties in computing at a constant order in the expansion due
to the non-renormalizable nature of chiral soliton models. Moreover,  there are fundamental difficulties in computing at a
constant order in the expansion due to the non-renormalizable
nature of chiral soliton models and the absence of any viable power counting scheme.  We show that the continuum mode contribution to the strangeness diverges, and as a result the computation of the strangeness content at leading non-vanishing order is not a well-posed mathematical problem in these models.

\end{abstract}

\maketitle

The extraction of the ``strangeness content of the nucleon'' has
been the subject of considerable experimental activity over the
past twenty years, largely involving parity violating electron
scattering\cite{exp}.  One class of models studied
extensively\cite{Kopeliovich:1997pq,Weigel:1997re,Wakamatsu:1995tv,Christov:1995vm,Ellis:1995de,Weigel:1995cz,Kim:1995hu,Kroll:1994um,Weigel:1992pm,Karliner:1991md,Park:1991fb,Praszalowicz:1989xw,Klabucar:1989si,Park:1989nz,Weigel:1989iw,Park:1989wz,Bernard:1988jf,Khatsimovsky:1987bb,Brodsky:1988ip,Wong:1990sy}
for the purpose of understanding the strangeness content ({\it
i.e.} the matrix elements of strange quark bilinear operators in
the nucleon) were chiral soliton models such as the
Skyrmion\cite{Skyrme}. The semi-classical treatment of these
models based on collective quantization has provided a reasonably
good description of many non-strange properties of the
nucleon\cite{ZahedBrown}. It is interesting to consider how well
these models do in describing the strange matrix elements For
example, the strange magnetic moment in these models typically
comes out in the range $\mu_s \sim -0.5$ in units of the nuclear
magneton  with the strangeness radius of about $r^2_s \sim -0.1
{\rm fm}^2$ (with the exact values depending on the variant of the
model used).  The experimental values for these appear to be
consistent with zero. A recent fit\cite{Thomas} to the world's
data yields $\mu_s=0.12 \pm 0.55$ and $r^s_s=0.01 \pm 0.95 {\rm
fm}^2$.  Thus the values predicted in the models appear to be on
the large size but do not appear to be inconsistent with the data.

What is one to conclude from this situation?  Before this issue
can be addressed, it is necessary to understand how these
quantities were computed in these models. We note that these
calculations have been based directly on collective quantization,
or some variation of this such as the Yabu-Ando
method\cite{Yabu:1987hm} which reduces to collective quantization
in the $SU(3)$ flavor limit. As will be shown in this letter,
calculations of strange quark matrix elements in these models
using collective quantization are inconsistent from the
perspective of the semi-classical expansion, and hence should not
be regarded as true predictions from the models. Moreover, the
non-renormalizability of such models creates fundamental
difficulties intrinsic to any consistent description of the
strangeness content.

These models are only known to be
meaningful in the context of a semi-classical expansion.  All
treatments of the model begin with a classical solution which
determines the gross structure of the baryon.  To proceed in a systematic fashion, one
must assume that a semi-classical treatment is valid.  In
this context, it will be shown that:
\begin{enumerate}
\item Strange quark matrix elements of the nucleon necessarily
occur at a subleading order in a semi-classical expansion.\label{1}

\item  Certain subleading effects---those associated with the
collective zero modes---are automatically included in treatments that
use collective quantization. However, other contributions {\it
which also occur at the first nonvanishing order}---those from the
non-collective continuum modes---are not included. Since
these effects have been neglected in typical computations of the
strangeness content, the computations are inconsistent. \label{2}

 \item The computation of the strangeness content at leading
non-vanishing order starting from the  Lagrangian of the chiral
soliton model in not a well-posed mathematical problem.
Contributions from the continuum modes contribute at the leading
non-vanishing order and are divergent. The models are not
systematic effective field theories with controlled power
counting; the divergences which appear cannot be absorbed by
renormalizing coefficients in the models as they are given in
their Lagrangians. \label{3}

\end{enumerate}

At first sight these results suggest a deficiency of Skyrme-type
models for these observables. However, they actually highlight a
strength: the ability of the models to encode the underlying
quark-loop structure of QCD.  The semi-classical expansion of the
models corresponds to an expansion in the number of closed quark
loops.   At the QCD level, strange quark matrix elements come from
quantum loops and thus have a qualitatively different origin than
do non-strange matrix elements of the nucleon. It is a virtue of
the semi-classical models that their structure forces one to
impose additional physics inputs in the form of new prescriptions
in order to describe the qualitatively distinct physics of the
strangeness content of the nucleon.

The framework of the analysis is the semi-classical expansion. One
natural way to justify it is via Witten's celebrated connection of
the semi-classical treatment of the soliton with the large $N_c$
limit of QCD \cite{WittenOrig}; the semi-classical expansion of the
soliton matches the $1/N_c$ expansion of QCD. One need not invoke
large $N_c$ as the ultimate justification of the semi-classical
expansion. However, regardless of how of the expansion is justified,
factors of $1/N_c$ may be used as markers to keep track of orders
in the semi-classical expansion. In the present context we work in
the semi-classical analysis analogous to the standard large $N_c$ limit
of 't Hooft\cite{'tHooft}, which is the appropriate limit when the usual ${\cal O}(N_c)$ strength
is taken for the WWZ term\cite{WittenGlobal}.

To illustrate the underlying issues in a relatively
simple context, we consider the strange scalar matrix element of
the nucleon in the chiral limit of the three flavor version of
Skyrme's original model\cite{Guadagnini:1983uv}. However, the conclusions depend
on neither the choice of observable or model. The action for the
model is
\begin{widetext}\be S=\int \, d^4 x \, \left ( \frac{f_{\pi}^2}{4}
\mathrm{Tr}(L_{\mu} L^{\mu})
    + \frac{\epsilon^2}{4} \mathrm{Tr}([L_{\mu},L_{\nu}]^2)
  + \frac{B_0}{4} \mathrm{Tr}\left( M (U+U^\dagger -2)\right )  \right ) +
    N_c S_{WWZ}
    \label{SK}
\ee
\end{widetext}
 where the left chiral current $L_\mu$ is given by
$L_{\mu} \equiv U^{\dagger}
\partial_\mu U$, with $U \in SU(3)_f$ \cite{Skyrme,ANW,WittenCurrent};
$S_{WWZ}$ is the Witten-Wess-Zumino (WWZ) term, whose inclusion is
necessary for the Skyrme model to correctly encode the anomaly
structure of QCD\cite{WittenGlobal, WittenCurrent}.  The $U$ field
can be written as $U = \exp \left ( i \sum_a \lambda_a \pi_a/f_\pi
\right )$ where the $\pi$ are the Goldstone boson meson fields and the
$\lambda_a $ are the Gell-Mann matrices;  $M$ is the quark mass matrix.
For simplicity the present analysis will be done in the chiral
limit; however, the mass term is included as an external source.
Thus the matrix element of interest---the strange scalar matrix
element in the chiral limit---can be obtained by computing the
nucleon mass for arbitrary values of the mass and then
differentiating:
 \be \langle
N| \overline{s} s - \langle \overline{s} s \rangle_{vac} |N\rangle
=  \frac{d M_N}{d m_s}  \label{mdif} \ee \;
\noindent ``vac''
indicates a vacuum value.

At the QCD level, the strangeness content of the nucleon can only
arise from quark loops.  This already establishes point \ref{1}
above: there is a suppression factor of $1/N_c$ for each quark
loop. Thus strange quark matrix elements are subleading in the
semi-classical expansion.

Consider the computation of the scalar strange quark matrix
element using collective quantization, as done originally by
Donahue and Nappi\cite{Nappi}. The computation simplifies somewhat
if one considers the ratio of the strange scalar matrix element to
the total scalar matrix elements of the three light flavors;
denote this ratio $X_s$:
\begin{equation}
X_s \equiv \frac{ \left \langle N \left | \overline{s} s -\langle
\overline{s} s \rangle_{\rm vac}  \right |N \right \rangle} {
\left \langle N \left | \overline{u} u  + \overline{d} d +
\overline{s} s -\langle \overline{u} u+ \overline{d} d +
\overline{s} s \rangle_{\rm vac}  \right |N \right \rangle} \; \;
\label{X}
\end{equation}
$| N \rangle$ is the nucleon state. In collective quantization,
the collective SU(3) rotation variables, $A$,  act on the standard
classical static hedgehog: $U=A^\dagger U_h A$ with the hedgehog
Skyrmion defined as $U_h \equiv \exp \left (i \hat{r} \cdot
\vec{\tau} f(r) \right)$ (where $\vec{\tau}$ are the first three
Gell-Mann matrices) ; $f(r)$ is the standard Skyrme profile
function for states of baryon number $B=1$. Evaluating $X_s$ using
standard SU(3) collective quantization\cite{Schechter:1999hg}
yields \be X_s = \frac{1}{3}\langle N \left| 1- D_{8 8}\right |
N\rangle = \frac{1}{3}\int dA \; \psi^*_{N}(A)\left(1 - D_{8 8}
\right ) \psi_{N}(A) \label{strangeElement} \ee \
\noindent where
$dA$ stands for the Haar measure on SU(3), $D_{8 8}=\frac{1}{2}
\textrm{Tr}\left[ \lambda_8 A \lambda_8 A^{\dag}\right]$ (which is
an SU(3) Wigner D-matrix), and $\psi_{N}(A)$ is the collective
wave function for the nucleon---{\it i.e.,} an appropriately
normalized SU(3) Wigner D-matrix.  Evaluating the expression using
the collective wave function for the nucleon
gives\cite{Nappi,Schechter:1999hg} $X_s=7/30 \approx .23$. While
this is relatively small numerically, it is non-zero.  Note that
this ratio does not depend on the form of the Skyrme profile
function $f(r)$.  This might suggest that it is a
model-independent result, but as pointed out by Kaplan and
Klebanov\cite{Kaplan:1989fc}, this simple result is not universal
and depends on the form of the mass term in the Skyrme lagrangian.

Apparently the standard leading order collective quantization
calculation used in the computation of strange quark matrix
elements includes subleading effects in the semi-classical
expansion.  As was noted in ref.~\cite{Cohenpenta} one must
include an explicit coefficient of $N_c$ rather than three as the
coefficient of the WWZ term in order to make explicit the counting
and trace orders in the semi-classical expansion. The coefficient
of the WWZ term constrains the allowed $SU(3)$ multiplets.  Thus at arbitrary $N_c$ the
nucleon is in the generalized representation
``$8$'', specified by $(p,q) = \left ( 1,\frac{N_c-1}{2} \right
)$\cite{Cohenpenta}. $X_s$ can be computed at arbitrary
$N_c$\cite{Cherman:2006xa} from Eq.~(\ref{strangeElement}) using standard
group theoretical methods and the use of SU(3) Clebsch-Gordan
coefficients appropriate for the ``$8$''
representation\cite{CohenLebed}. The result\cite{Kaplan:1989fc,Kopeliovich:2005hs,Cherman:2006xa} is
\begin{eqnarray}
&{}&\left \langle N \left | \overline{s} s  - \langle \overline{s}
s \rangle_{\rm vac}  \right |N \right \rangle^{\rm coll.  quant.}  \\
& = & \frac{2( N_c +4)}{N_c^2 + 10 N_c +21} \left \langle N \left
| \overline{u} u  + \overline{d} d + \overline{s} s -\langle \overline{u} u+
\overline{d} d + \overline{s} s\rangle_{\rm vac} \right |N \right \rangle
\nonumber \\ & = & \left ( \frac{2}{N_c}  +{\cal
O}\left(1/N_c^{2}\right) \right)\left \langle N \left |
\overline{u} u + \overline{d} d  -\langle \overline{u} u+
\overline{d} d + \rangle_{\rm vac} \right |N \right \rangle \;
\nonumber . \label{Xn}
\end{eqnarray}
Clearly, $\left \langle N \left | \overline{s} s  - \langle
\overline{s} s \rangle_{\rm vac}  \right |N \right \rangle  $ is
subleading in $N_c$---and, hence, in the semi-classical
expansion---as compared to its non-strange analog.

The collective quantization method builds in {\it some}
contributions to $ \langle N| \overline{s} s - \langle
\overline{s} s \rangle_{vac} |N \rangle$ at the leading
nonvanishing order in the semi-classical expansion ({\it i.e.},
the first subleading order). The question is whether it captures
all of them.  The answer is no: there are  contributions to
$\langle N| \overline{s} s - \langle \overline{s} s \rangle_{vac}
|N\rangle$ at first subleading order which are not included in the
collective quantization. These may be computed via
Eq.~(\ref{mdif}): one differentiates the first subleading
contribution to the nucleon mass and with respect to $m_s$. The
procedure for implementing the semi-classical expansion for the
calculation of the mass of a topological soliton in a bosonic
theory is very well
established\cite{Callan:1985hy,Klebanov:1989iy,Itzhaki:2003nr}:
The boson fields are expanded around the classical solution to
quadradic order and and then quantized. The next-to-leading
contribution to the mass is simply the energy of the zero point
motion of these harmonic modes.

In general there are contributions from both discrete eigenmodes
and from continuum modes.  The modes in the SU(3) Skyrme model
around the standard hedgehog can be broken up into kaon modes and
pion modes.  From the structure of Eq.~(\ref{SK}), it is clear
that only the kaon modes depend on $m_s$ and contribute to the strangeness content. The kaon modes separate
into modes carrying strangeness plus or minus one,
corresponding to kaons and anti-kaons. Moreover, eigenmodes carry good
total orbital angular momentum $L^2$ and good ``grand spin''
$\vec{g}= \vec{I} +\vec{L}$ with $g=L \pm 1/2$ \cite{Callan:1985hy,Klebanov:1989iy}. Thus,
the contribution to the strangeness content at next-to-leading
order ({\it i.e.} leading non-vanishing) in the semi-classical
expansion is:
 \begin{eqnarray} &{}& \langle
N| \overline{s} s - \langle \overline{s} s \rangle_{vac}
|N\rangle_{\rm NLO}=   \\
&{}&\frac{1}{2}    \sum_{g,L} (2g+1) \, \, \frac{d}{d m_s} \left (
\sum_n \omega_{L g +; n} ^{\rm disc} \, + \, \sum_n \omega_{L g
-;  n} ^{\rm disc}  \right ) \, + \nonumber \\
  &{}&\frac{1}{2} \, \sum_{g,L} (2g+1) \, \,  \frac{d}{d m_s}
\int \frac{d \, \omega}{\pi} \left ( \delta'_{L g +}(\omega ) +
\delta'_{L g -}(\omega) \, \right) \omega  \nonumber \label{mdif2}
\end{eqnarray}
where  $\delta_{L g \pm}$ is the phase shift for given $L$, g-spin
and strangeness $\pm 1$ and  $\omega_{L g \pm;  n}^{\rm disc}$
indicates the $n^{\rm th}$ discrete frequency with fixed grand
spin and strangeness.

The frequency of the discrete modes and the phase-shifts for the
continuous modes can be computed from the equations of motion for
the kaon fluctuation around the soliton may be derived in the
manner of Callan and Klebanov~\cite{Callan:1985hy,Klebanov:1989iy}.  A compact form for
these is given in ref.~\cite{Klebanov:1989iy}.  The equations are naturally
expressed in terms of dimensionless lengths and masses:
$\tilde{r} = \frac{f_\pi}{\epsilon} r$, $\tilde{\omega} =
\frac{\epsilon}{f_\pi} \omega$ and $\tilde{m}_K=
\frac{\epsilon}{f_\pi} m_K$.  (Note that the conventions used here
differ from ref.~\cite{Klebanov:1989iy}: the symbol $f_{\pi}$ here corresponds to
$f_{\pi}/2$ in ref.~\cite{Klebanov:1989iy} and $\epsilon$  corresponds
$\frac{1}{2 \sqrt{2} e}$).  The equations for the modes are
\begin{equation}
\left (y(\tilde{r}) \, \tilde{\omega}^2 \mp 2  \lambda(\tilde{r})
 \, \tilde{\omega} + \Theta \right )k^{\tilde{\omega}}_{l,g,\pm
}(\tilde{r})  =  0 \label{mode}
\end{equation}
with
\begin{eqnarray}
\Theta & \equiv & \tilde{r}^{-2} \partial_{\tilde{r}}
h(\tilde{r}) \partial_{\tilde{r}} - \tilde{m}_K^2 -V_{\rm
eff}(\tilde{r}) \, , \; \; \; \;\lambda(\tilde{r})\equiv
-\frac{N_c
f'\sin(f)}{8 \pi^2 \epsilon^2 \tilde{r}^2} \, ,\nonumber \\
y(\tilde{r}) &\equiv&   1 + 2 s(\tilde{r}) + d(\tilde{r}) \, ,
\; \; \; \; h(\tilde{r}) \equiv   1 + 2 s(\tilde{r})  \, , \; \nonumber \\
d(\tilde{r}) & \equiv& {f'}^2 \, , \; \; \; \; s(\tilde{r})
\equiv \sin^2(f)/\tilde{r}^2 \, ,\; \; \; \;c(\tilde{r})\equiv
\sin^2\left({f}/{2} \right ) \nonumber
\end{eqnarray}
where $f$ is the Skyrme profile, the prime indicates
differentiation with respect to $\tilde{r}$ and
\begin{eqnarray}
V_{\rm eff} & = & - \frac{d+ 2 s}{4} -2 s(s +2 d) \nonumber \\ &
+ & \frac{(1+d+s) (L(L+1)+2c^2 +4 c \vec{I}\cdot \vec{L}
)}{\tilde{r}^2} \nonumber \\ & +& \frac{6}{\tilde{r}^2}\left(  s (c^2 + (2
c-1)\vec{I}\cdot \vec{L}) +\partial_{\tilde{r}} \left ( (c+
\vec{I}\cdot \vec{L}) f' \sin (f)   \right)\right) \; .
\nonumber
\end{eqnarray}
In Eq.~(\ref{mode}), the $\mp$ indicates the strangeness of the
mode, $g$ the g-spin and $L$ the orbital angular momentum.  Phase
shifts may be extracted from the modes by comparing with solution
for $f=0$.

It is known that there is only one discrete mode for this system
\cite{Itzhaki:2003nr}. The mode has $L=1$, $g=1/2$ and $s=-1$. Moreover, at
$m_s=0$ the mode is collective and associated with flavor
rotations out of the SU(2) subspace of the original hedgehog:
$k^{\rm disc}_{L g s}(r) =k^{\rm disc}_{1 \, \frac{1}{2} \, -}(r)
= \sin \left ( f(r)/2 \right )$ where $k^{\rm disc}_{1 \,
\frac{1}{2} \, -}(r)$ is the spatial profile of the mode.  By
construction, at $m_s=0$ this collective mode is in a flat
direction and has zero frequency. It makes a nonzero contribution
in Eq.~(\ref{mdif2}), however, since the derivative of the
frequency with respect to $m_s$ is nonzero at $m_s=0$. The
frequency of this mode can be expanded perturbatively in $m_K^2$
from the underlying equations of ref.~\cite{Klebanov:1989iy}; one finds
$\omega^{\rm disc}=\frac{4 \, m_K^2 \, f_\pi^2}{N_c} \int  d^3 x
\left (1 - \cos \left ( f \right) \right )
 $ which in turn implies that the discrete mode contribution to
the strangeness content at first subleading order in the
semi-classical expansion is:
\begin{equation}
\langle N \left | \overline{s} s -\langle \overline{s} s
\rangle_{\rm vac} \right | N \rangle ^{\rm disc} = \frac{2}{N_c}
\langle N \left | \overline{u} u + \overline{d} d -\langle
\overline{u} u + \overline{d} d\rangle_{\rm vac} \right | N
\rangle \; . \label{Xd} \end{equation}

As expected, the contribution from the discrete mode in
Eq.~(\ref{Xd})---associated with the collective motion in the flat
direction---exactly reproduces the result of collective
quantization given in Eq.~(\ref{Xn}) at first nonvanishing order
in the semiclassical  ($1/N_c$) expansion. However, the collective
quantization does {\it not} include the contributions from  the
continuum modes. These contributions are clearly both nonzero
generically---the phase shifts are nonzero and dependent on $m_s$
\cite{Klebanov:1989iy}---and are of the same order in the
semi-classical expansion as the discrete mode contribution encoded
in the collective quantization. Thus, calculations of the
strangeness content which neglect these are unjustified from the
perspective of the semi-classical expansion. Since these {\it are}
neglected by those calculations on the market which purport to
compute the strangeness in chiral
solitons\cite{Kopeliovich:1997pq,Weigel:1997re,Wakamatsu:1995tv,Christov:1995vm,Ellis:1995de,Weigel:1995cz,Kim:1995hu,Kroll:1994um,Weigel:1992pm,Karliner:1991md,Park:1991fb,Praszalowicz:1989xw,Klabucar:1989si,Park:1989nz,Weigel:1989iw,Park:1989wz,Bernard:1988jf,Khatsimovsky:1987bb,Brodsky:1988ip,Wong:1990sy},
one must regard these calculations as being inconsistent with the
semi-classical expansion\footnote{Ref. \cite{Wong:1990sy} is an
exception and does discuss the mode sums.  However it does so in
the context of the kaon number, an unphysical quantity, which is
related to the strangeness content via an uncontrolled
approximation.}. This establishes point \ref{2}. We note that this
point is implicit in  the work of Kaplan and
Klebanov\cite{Kaplan:1989fc}.

The cure for this problem seems obvious: one ought to simply
include these continuum mode contributions in the calculation.
Unfortunately, the contribution from these modes  diverges.

This divergence can be seen from the form of Eq.~(\ref{mode}). For
large $\omega$ and fixed $L$ the  system is in the WKB regime. At
$L=0$ and high $\omega$, a simple WKB calculation at lowest order
in $\omega^{-1}$ yields that
\begin{eqnarray}
\left. \frac{\partial \delta}{\partial m_s}\right |_{(m_s=0)} & =
&
\frac{\alpha}{{\omega}} + {\cal O}({\omega}^{-3})  \label{WKB} \\
{\rm with} \; \; \; \; \alpha &=& \frac{\epsilon}{f_\pi}  \frac{d
m_K^2}{d m_s} \, \int{d\tilde{r}} \; \frac{\tilde{r}}{2 \sqrt{
h(\tilde{r})\, y(\tilde{r}) } } \; .\nonumber
\end{eqnarray}
In fact, Eq.~(\ref{WKB}) holds for all partial waves (although the
value of $\omega$ at which  the asymptotic regime sets in  grows
with $L$).  The reason for this is that although the angular
momentum potential barrier grows with $L$, at any fixed $L$ we are
free to choose an arbitrarily high $\omega$.  This allows the WKB
region to penetrate close to the origin, where the sum of the in-
and out-going wave functions must vanish as a boundary condition,
just as it does for $L=0$.  Recalling Eq.~(\ref{mode})  one sees
from the above that the continuum mode sum contributions in all
channels diverge logarithmically in the ultraviolet limit in a
universal way. Thus the continuum mode contribution to the
strangeness content of the nucleon is divergent. This divergence
is not surprising: it reflects the one-loop divergences present in
the underlying mesonic theory.

To make meaningful predictions from the theory, one must render
this divergence finite in a manner consistent with the theory. If
the theory were renormalizable this would be a well-defined task;
any divergence which arises in the loops could be absorbed by
renormalization of the constants in the original theory. However,
chiral soliton models such as the Skyrme model are {\it not }
renormalizable.  More significantly, chiral soliton models are not
effective field theories since they lack a systematic power
counting scheme. Terms with any number of derivatives of the meson
fields contribute at every order in the $1/N_c$ expansion. Thus
one cannot use power counting to restrict  the  number and type of
counterterms at next-to-leading order in the semi-classical
expansion: one needs an entirely {\it ad hoc} and uncontrolled
prescription.

Of course, the act of building a chiral soliton model in the first
place required making a similarly bold prescription: of the
infinite number of terms which could be included at leading order
in the $1/N_c$ expansion only a very few are kept.  The troubling
issue here, however, is that unlike for leading order observables,
the initial prescription used to set up the model is not
sufficient to compute strange quark matrix elements; an additional
prescription is needed. Since the computation of all strange quark
matrix elements completely depends on the prescription used at
next-to-leading order, the initial model given by the lagrangian,
on its own, has no predictive power for these matrix elements.

This implies that the problem of computing the strangeness content
from chiral soliton models at the first nonvanishing order in the
semi-classical approximation is not well-posed. To proceed, one
must make some prescription not fixed from the Lagrangian of the
original model. This is highly problematic in that result for the
strange content is not fixed by the original model. This
establishes our final point. As we noted above,  this is
unsurprising: the strangeness content arises from quark loops.
This is qualitatively different from the dominant origin of
non-strange matrix elements and thus new physical inputs are
required.

It should be clear that the conclusions in this letter are quite
general. Although we have focused on the problem of computing the
scalar matrix element at zero momentum transfer at the chiral
limit of the Skyrme model, the structure of the argument holds
quite generally. The argument that the contributions come from
quark loops and must be subleading in the semi-classical expansion
holds generally for any strange operator in any model and
regardless of whether the system is in the chiral limit. We have
shown this explicitly for the scalar matrix elements in the Skyrme
model by demonstrating that the collective quantization leads to
contributions which are subleading in $1/N_c$ (and hence in the
semi-classical expansion). We have explicitly verified that the
same thing occurs for the case of the strange electric form
factor---as it must. The general argument that the quantum
fluctuations of all modes, collective and non-collective alike,
contribute at the lowest nonvanishing order in the semi-classical
expansion again holds for any strange matrix element in any model
whether in the chiral limit or not. The need for a prescription
not contained in the original model to compute at the leading
nonvanishing order in the semi-classical expansion also applies to
all strange quark observables in any non-renormalizable chiral
soliton model.

\acknowledgments This work is supported by the U. S. Department of Energy under
grant number DE-FG02-93ER-40762. We are grateful to David Kaplan, Igor Klebanov, and Victor Kopeliovich for helpful discussions.

\end{document}